%Paper: hep-th/9301118
%From: Ofer Aharony <OFERA@taunivm.tau.ac.il>
%Date: Thu, 28 Jan 93 15:57:59 IST
%Date (revised): Sun, 31 Jan 93 12:55:27 IST

\input phyzzx
% Fusion rules paper (January 1993)

%macropackage=phyzzx
\overfullrule=0pt
\def\nijk{N_{ij}{}^k}
\def\xt{\tilde x}
\def\yt{\tilde y}
\def\xy{(x+y)}
%*********************************************
\def\cmp#1{{\it Comm. Math. Phys.} {\bf #1}}

\def\np#1{{\it Nucl. Phys.} {\bf B#1}}

\def\jmath#1{{\it J. Math. Phys.} {\bf #1}}
\def\mpl#1{{\it Mod. Phys. Lett.}{\bf A#1}}

\REF\DG{D. Gepner, \cmp {141} 381-411 (1991) }
\REF\DFZ{P. Di Francesco and J.-B. Zuber, "Fusion potentials I",
SPhT 92/138, hep-th/9211138}
\REF\EV{E. Verlinde, \np {300} [FS22] 360-376 (1988) }
\REF\GS{D. Gepner and A. Schwimmer, \np {380} 147-167 (1992) }
\REF\BMRS{M. Bourdeau, E.J. Mlawer, H. Riggs, H.J. Schnitzer, \mpl {7}
689-700 (1992) }

%************************************************************************
%************************************************************************
%title page
\rightline{TAUP-2029-93}
\rightline{hep-th/9301118}
\date{January 1993}
\titlepage
\vskip 1cm
\title{ Generalized fusion potentials}
\author {Ofer Aharony \footnote{\dagger}{Work supported in part by
the US-Israel Binational Science Foundation and the Israel Academy
of Sciences.}}
\address{ School of Physics and Astronomy\break
Beverly and Raymond Sackler \break
Faculty of Exact Sciences\break
Tel Aviv University\break
Ramat Aviv, Tel-Aviv, 69978, Israel}

\abstract{
Recently, DiFrancesco and Zuber have characterized the RCFTs which have
a description in terms of a fusion potential in one variable, and
proposed a generalized potential to describe other theories. In this
note  we give a simple criterion to determine when such a generalized
description is possible. We also determine which RCFTs can be described
by a fusion potential in more than one variable, finding that in fact
all RCFTs can be described in such a way,
as conjectured by Gepner.}
\endpage

\chapter{ Introduction}
 Rational conformal field theories (RCFT's) are characterized by the
fusion rules of their (finite number of) fields :
$$ \phi_i \phi_j = \sum_k \nijk \phi_k \eqn\fusion $$
where the indices $i,j,k$ run over the $N$ primary chiral fields in the
operator algebra of the RCFT, and the $\nijk$ are non-negative integers.
 This algebra is commutative and associative, giving symmetry
and crossing constraints on the structure constants $\nijk$.
As shown by E. Verlinde  \refmark\EV , the fusion
rules can be expressed in terms of the unitary modular matrix $S$
of the RCFT :
$$ \nijk = \sum_l {{S_{il}S_{jl}S_{kl}^*} \over S_{1l}} \eqn\nijks $$
where $\phi_1$ is the identity operator.

   A trivial representation of this algebra via $N \times N$ matrices
can be obtained by representing the field $\phi_i$ as the matrix
$$ {(\phi_i)}_{jk} = \nijk . \eqn\repmat $$
These matrices form a representation
of the fusion algebra as can be checked using the associativity
conditions. It follows from \nijks\ that
the matrix $S$ diagonalizes all the
matrices $\phi_i$, and their eigenvalues are of the form :
$$ \lambda_i^{(l)} = {S_{il} \over S_{1l}} \eqn\eigenvals $$
where $\lambda_i^{(l)}$ designates the $l$`th eigenvalue of $\phi_i$.

It was shown by Gepner \refmark\DG\ that any such algebra
can be represented as a ring of polynomials in variables $x_1,...,x_n$
modulo some ideal, such
that the multiplication of fields becomes simple polynomial
multiplication modulo several polynomials. Moreover,
Gepner conjectured \refmark\DG
that this ideal of constraints can always be derived from
a potential, and has explicitly
shown that this is true for the fusion rules of
$SU(N)_k$ (this has since been shown for other Kac-Moody
theories as well \refmark\BMRS \refmark\GS ).
Such a representation is useful because it has a simple geometrical
interpretation \refmark\DG\ as well as a physical interpretation (such
as in Landau-Ginzburg models). The demands we have from such a
representation is that the trivial polynomials $x_i$ will
represent actual primary
fields of   the theory (for $i=1,...,n$), that the representatives of
different fields will be linearly independent, and of course that the
correct fusion rules will be recovered from the ring of polynomials
modulo the derivatives of this potential. To recover the fusion rules
of the RCFT from such a representation we need to know the way all
primary fields are represented as polynomials in addition to knowing
the potential.

 Di Francesco and Zuber \refmark\DFZ\ have examined which theories
have a representation in terms of a fusion potential in one variable.
They found a simple criterion to determine whether this is possible : it
is possible whenever the field we choose to be represented by $x$ has no
degenerate eigenvalues (otherwise there is no polynomial representation
at all, let alone a potential representation).
They have also proposed a generalized representation in which one does
not demand that all fields are linearly independent over
the real numbers ${\cal R}$,
but only that they are linearly independent over
the rational numbers ${\cal Q}$.  This is
enough to enable recovering the fields' fusion rings from the polynomial
ring. In ref. [\DFZ]    several examples of such representations were
given.  In section 2 of this note we will give a simple criterion
determining whether such a representation is possible or not (and in
fact providing a simple way to construct all such representations
whenever they exist).

In section 3, we examine fusion potentials using polynomials in more than
one variable. We find that the generalization of the results of
ref. [\DFZ]  to this case is quite straightforward, and that whenever we
have fields $x_1,...,x_n$ such that no two eigenvectors are degenerate
for all of them, we can represent the theory by a potential in these
variables. This will prove Gepner's conjecture, since in any RCFT there
are no eigenvectors which are degenerate for all fields. Unfortunately,
we will find that there exist many different potential descriptions for
a single RCFT, meaning that the potential of a theory is probably not
the best way to characterize it.

\chapter{ Generalized one-variable fusion potentials}
As mentioned above, all the matrices representing the fields in their
matrix representation have the same eigenvectors, which we shall denote
below
by $v_j$ (for $j=1,...,N$). We can look at these vectors as combinations
of fields, in which case they are given by
$$ v_j = \sum_i S_{ij}^* \phi_i  \eqn\eigenvect $$
or equivalently
$$ \phi_i = \sum_j S_{ij} v_j.  \eqn\noteigenvect $$
By definition the eigenvectors satisfy
$$ \phi_iv_j = \lambda_i^{(j)}v_j  \eqn\phiv $$ where $\lambda_i^{(j)}$
are given by \eigenvals, and from these equations one can easily check
that fusion of two eigenvectors gives
$$ v_iv_j = \delta_{ij}v_j {\lambda_i^{(j)} \over S_{ij} }  \eqn\vivj $$
or
$$ v_iv_j = \delta_{ij} v_j {1 \over S_{1j}}. \eqn\morevivj $$

In this section we will be interested in representations of the fusion
algebra in terms of a fusion potential in one variable.  It is obvious
that in this case we get just one polynomial constraint on our variable
$x$ (${\partial V(x) \over \partial x} = 0$) and that any such constraint
can be derived from a potential (by integration). From the matrix
representation described above, it is clear that the constraint of
minimal degree
satisfied by a field $x$ is exactly the minimal polynomial of the matrix
that represents it. In [\DFZ] it was shown that when this polynomial is
of degree $N$ (so that all eigenvalues of $x$ are different) we can
represent the fusion algebra as the algebra of polynomials in $x$ modulo
the constraint given by this minimal polynomial (which can obviously
be derived from a potential).  If the eigenvalues of $x$ are denoted
by $\mu_j$ the represention of $\phi_i$ is given simply by the
polynomial of degree $N-1$ that transforms $\mu_j$ to $\lambda_i^{(j)}$
for $j=1,...,N$.
Obviously, if we do not demand that $x$ be a field from the theory, we
can always build such a representation in terms of any $x$ which has
no degenerate eigenvalues, but usually we want $x$ to be one of our
fields and in
[\DFZ] it was shown that this is only possible when $x$ has no
degenerate eigenvalues. Theories which have no field without degenerate
eigenvalues (such as the unitary
minimal models excluding the Ising model, or general $(p,q)$ minimal
models with $p,q > 3$)
cannot be, therefore,
represented by a fusion potential in one variable, in the usual way.

In the next section we will
show that such theories can always be represented
by a potential in more than one variable.     For now let us stay with
the one variable case and, following [\DFZ], try to relax some of our
requirements from the representation to
enable such a description after all.
Let us look at a certain field $x$ from which we wish to generate our
polynomial algebra.  If $x$ has $m$ different eigenvalues, it's
minimal polynomial is of degree $m$. Hence, we can only have $m$
different polynomials which are not linearly dependent over {\cal R}
(for instance, we can take them to be $1,x,...,x^{m-1}$).
However, to reconstruct the fusion algebra from the
potential, all we need is that the polynomials representing the different
fields be linearly independent over {\cal Q}, since we know that the
coefficients $\nijk$ are all integers. In [\DFZ], Di Francesco and Zuber
 have analyzed some
examples for which such a generalized representation is possible and
some for which it is not. We shall now analyze the general case and
construct all possible representations of this form
for a given theory.

Suppose we wish to know if we can find such a
representation for a given field $x$ with eigenvalues $\lambda_i$,
(for $i=1,...,m$)
such that the minimal polynomial of $x$ (which obviously still must be
the constraint we impose on the polynomial algebra) is given by
$$ V'(x) = \prod_{i=1}^m(x-\lambda_i).  \eqn\vprime $$
We are looking for polynomial representations of all other fields
$\phi_i$ which will satisfy the fusion algebra (modulo $V'(x)$) and
which will be linearly independent over ${\cal Q}$. Since there is
a non-singular linear transformation between the fields and the
eigenvectors (given by \eigenvect,\noteigenvect), this is equivalent
to finding a polynomial representation for the eigenvectors $v_j$,
which should satisfy the same requirements. Suppose that $v_i$ is
an eigenvector with an eigenvalue $\lambda_k$ for $x$ : in this case
one of the equations which should be satisfied modulo $V'(x)$ is
$xv_i = \lambda_kv_i$ or $(x-\lambda_k)v_i = 0$.  From the form
of $V'(x)$ \vprime\ it is clear that the only possible form for
the polynomial representing $v_i$ (to be denoted by $V_i$) is
$$ V_i(x) = \alpha_i \prod_{j\neq k}
{x-\lambda_j \over \lambda_k-\lambda_j }  \eqn\polyv $$
where the product goes over all different eigenvalues of $x$, and
$\alpha_i$ is a normalization constant (chosen so that
$V_i(\lambda_k) = \alpha_i $). Now let us use equation \morevivj\
for $i=j$.  Since the equation should be satisfied exactly at the
point $x=\lambda_k$ where $V'(x)$ vanishes, we find that
$$ \alpha_i^2 = {1 \over S_{1i}} \alpha_i. \eqn\eqalpha $$
Thus, for all $i$, either $\alpha_i = 0$ or
$\alpha_i = {1 \over S_{1i}} $. For all such choices equation
\morevivj\ is trivially satisfied if $v_i$ and $v_j$ correspond to
different eigenvalues of $x$. However, if it is to be satisfied for
two eigenvectors $v_i$ and $v_j$ corresponding to the same eigenvalue it is
necessary that either $\alpha_i$ or $\alpha_j$ be zero. This proves that
at most one of the $\alpha_i$ corresponding to some eigenvalue
$\lambda_k$ of $x$
can be different than zero. In fact, exactly one must be different from
zero.  This can be seen from the equation
$$ 1 = \sum_j S_{1j} V_j(x)  \eqn\one $$ which is equation \noteigenvect
\ taken for $i=1$, and which must be satisfied exactly for $x=\lambda_k$.
We have thus found which        polynomial representations in terms of
a field $x$ we can build in a certain theory (without yet demanding
linear independence of the polynomials representing the fields).  For
every eigenvalue $\lambda_k$ of $x$ we should choose one eigenvector
which has this eigenvalue, set it's polynomial to be \polyv\ with
$\alpha_i={1 \over S_{1i}} $, and set the polynomials of all other
eigenvectors with the same eigenvalue to zero. From the polynomials of
the eigenvectors we get via equation
 \noteigenvect\ the polynomials representing
the fields. The number of different possible representations utilizing
a certain field $x$ is therefore the product of the multiplicities of
it's eigenvalues, and we have given a prescription for the construction
of all such representations.

Next we should check when this construction gives a faithful polynomial
representation, with fields
which are linearly independent over {\cal Q}.
We, therefore, check if there
exists a rational linear combination of the polynomials which is zero.
Since the transformation from the fields to the eigenvectors is
nonsingular, a linear combination of fields is equivalent to a linear
combination of eigenvectors, and from our construction it is clear that
such a combination is zero if and only if it contains only eigenvectors
which we have set to zero (there are $N-m$ such eigenvectors).
The question, therefore, is whether in the vector space generated by
the eigenvectors which we have set to zero there is a vector with
rational coefficients (in terms of the fields) or not. This can easily
be checked for a given theory, as demonstrated below for some examples,
but we have not been able to obtain a simpler criterion to determine
when this is possible without constructing all eigenvectors.

Let us now analyze two simple examples which were also analyzed in
[\DFZ] from this point of view. Let us start with the $D_{2\nu +2}$
which has been shown in [\DFZ] to have such a representation whenever
$2\nu +1$ is not a square of an integer. From the considerations above
this follows straightforwardly.
It turns out that the field $x=\phi_1$ has
exactly one degenerate eigenvalue, $\lambda=-1$, which has two
corresponding eigenvectors given (in the basis
$\phi_0,\phi_1,...,\phi_{\nu-1},\phi_{\nu}^+,\phi_{\nu}^-$ and in a
convenient normalization) by
$$ \eqalign
{v_1 &= (1,-1,1,...,(-1)^{\nu-1},{1\over 2}(1+{\sqrt{2\nu+1}}),
 {1\over 2}(1-{\sqrt{2\nu+1}}))\cr
v_2 &= (1,-1,1,...,(-1)^{\nu-1},{1\over 2}(1-{\sqrt{2\nu+1}}),
 {1\over 2}(1+{\sqrt{2\nu+1}}))\cr  } \eqn\dnuvects $$
A potential representation of the algebra in terms of $x$ is therefore
obtained by setting one of these eigenvectors to zero (and no other
eigenvector), and this vector is a rational combination of the fields
if and only if $2\nu +1$ is the square of an integer, as was
derived in [\DFZ]. For other $\nu$ we obtain in this way exactly the
two possible representations given in [\DFZ].

Our second example will be the $(4,5)$ minimal model (the tricritical
Ising model) for which several such representations were obtained in
[\DFZ]. The eigenvectors for this case are (written in the basis
$\phi_{(1,1)},\phi_{(1,2)},\phi_{(1,3)},\phi_{(1,4)},\phi_{(2,1)},
 \phi_{(2,2)}$ for the fields) :
$$ \eqalign {
v_1 &= (1,\ \mu_1,\mu_1,\ 1,\ \sqrt2,\ \sqrt2\mu_1)   \cr
v_2 &= (1,\ \mu_1,\mu_1,\ 1, -\sqrt2, -\sqrt2\mu_1)   \cr
v_3 &= (1, -\mu_1,\mu_1, -1,\      0,\           0)   \cr
v_4 &= (1,\ \mu_2,\mu_2,\ 1,\ \sqrt2,\ \sqrt2\mu_2)   \cr
v_5 &= (1,\ \mu_2,\mu_2,\ 1, -\sqrt2, -\sqrt2\mu_2)   \cr
v_6 &= (1, -\mu_2,\mu_2, -1,\      0,\           0)   \cr }
\eqn\eigenvmin $$ where
$\mu_{1,2}$ are the two roots of the equation $x^2-x-1=0$, and
the eigenvectors of the various fields are given
(in the above order for the eigenvectors) by :
$$ \eqalign {
\lambda_{(1,1)}^i &= (1,1,1,1,1,1)    \cr
\lambda_{(1,2)}^i &= (\mu_1,\mu_1,-\mu_1,\mu_2,\mu_2,-\mu_2)  \cr
\lambda_{(1,3)}^i &= (\mu_1,\mu_1,\mu_1,\mu_2,\mu_2,\mu_2)  \cr
\lambda_{(1,4)}^i &= (1,1,-1,1,1,-1)    \cr
\lambda_{(2,1)}^i &= (\sqrt2, -\sqrt2, 0, \sqrt2, -\sqrt2, 0)  \cr
\lambda_{(2,2)}^i &= (\sqrt2\mu_1, -\sqrt2\mu_1, 0,
\sqrt2\mu_2, -\sqrt2\mu_2, 0)  \cr } \eqn\eigenvalmin  $$
We can see that any field (except $\phi_{(1,1)}$) can be chosen as $x$.
This follows from the observation that
the only combinations of eigenvectors which
have rational coefficients in terms of the fields include $v_3$ and
$v_6$ or $v_1$,$v_2$,$v_4$ and $v_5$. For example, if we wish $x$ to
be $\phi_{(1,1)}$ we have 4 possibilities, since
 either $v_1$ or $v_2$ and
either $v_4$ or $v_5$ must be set to zero, giving exactly the 4
representations given in [\DFZ] for this case. For $x=\phi_{(2,1)}$ we
have 8 different representations which are also all faithful. For
$x=\phi_{(2,2)}$ we must set to zero $v_3$ or $v_6$, and we obtain
a representation with a potential of the highest possible degree.
The potential of the lowest possible degree is obtained if we wish to
take $x=\phi_{(1,4)}$.  In this case only one of $v_1$,$v_2$,$v_4$,$v_5$
is different from zero, and only one of $v_3$,$v_6$, giving altogether
8 possible
representations. For example, if we choose $v_1$ and $v_6$ to be non-zero
we obtain the representation :
$$ \eqalign {
\phi_{(1,1)} &= 1   \cr
\phi_{(1,2)} &= {1 \over 2} x + {{\sqrt 5} \over 2}   \cr
\phi_{(1,3)} &= {{\sqrt 5} \over 2} x + {1 \over 2}    \cr
\phi_{(1,4)} &= x  \cr
\phi_{(2,1)} &= {1 \over \sqrt2} (x + 1)    \cr
\phi_{(2,2)} &= {\mu_1 \over \sqrt2} (x + 1)  \cr } \eqn\fields $$
which satisfies the algebra when taken modulo the constraint
$V'(x) = x^2 - 1 = 0$.

\chapter{ Multi-variable fusion potentials}
As Gepner has shown in ref. [\DG] , any RCFT can be represented as a ring
 of polynomials modulo some ideal of polynomials. This ideal is exactly
the ideal of polynomials which vanish at all of the points
$(\lambda_1^{(i)},\lambda_2^{(i)},...,\lambda_n^{(i)})$
where we have chosen the polynomials to be polynomials in variables
 $x_1,...,x_n$ corresponding to the fields $\phi_1,...,\phi_n$, and
where $i$ goes over the $N$ eigenvectors of the theory.  We wish to find
 a potential $V(x_1,...,x_n)$ whose derivatives will generate this ideal,
meaning that any function vanishing at all of the above points can be
 written as a (polynomial) linear combination of the derivatives of the
potential $V$. Since all the fusion rules are exactly satisfied at the
above points (as seen from the matrix representation of the algebra),
 they will all be generated by the potential and vice versa.
Obviously this is only possible when all of the above points are
different (since if all of the fields $\phi_i$ have degenerate
eigenvalues for some pair of eigenvectors, there is no way to represent
a field which is not degenerate for these eigenvectors as a polynomial in
 them). This is a necessary condition, and we will show that it is also
sufficient for the existence of a polynomial representation.

We will start by analyzing the simple case of a theory represented by
polynomials in two fields, which we shall denote by $x$ and $y$, and
where $x$ has no degenerate eigenvalues.  Of course in this case there
is a representation of the theory in terms of polynomials in $x$ alone
\refmark\DFZ, but we will later be able to generalize to other cases.
Let us denote the aforementioned points by $(\lambda_i,\mu_i)$ :
we will look for a potential of the form $V(x,y) = P(x) + yQ(x)$ which
gives constraints of the form
$$\eqalign { Q(x) &= 0 \cr P'(x) + yQ'(x) &= 0 \cr }  \eqn\constraint$$
which must be satisfied only at the above points. This can be trivially
solved by choosing $Q(x)$ to be a polynomial vanishing only at the
points $\lambda_i$, namely $Q(x) = \prod_i (x - \lambda_i)$, and by
choosing $P(x)$ to satisfy the $N$ equations
$$P'(\lambda_i) + \mu_i Q'(\lambda_i) = 0  \eqn\eqforp$$ for $i=1,...,N$.
For example, we could choose $P(x)$ to be the polynomial of the lowest
degree satisfying these constraints, which is
$$P(x) = \sum_i (-\mu_i) \int_{x_0}^x dx' \prod_{j\neq i} (x'-\lambda_j)
 \eqn\px $$ for any $x_0$.
 For this choice of the potential, it is obvious that the points in
which it's derivatives vanish are exactly the desired points, and we will
show explicitly that the algebra satisfied by the matrices $x,y$ can
indeed be represented as the algebra of polynomials modulo the
derivatives of this potential.

In general, the polynomials in this algebra are linear combinations of
the polynomials $x^ny^m$ for all $n,m$, and only $N$ of these polynomials
(viewed as matrix polynomials) are linearly independent.  Since we chose
$x$ not to have degenerate eigenvalues these can be chosen to be the
polynomials $1,x,x^2,...,x^{N-1}$. We need to show that all other
polynomials can be represented as linear combinations of these (modulo
the derivatives of the potential) and that they give rise to
the correct algebra.
For polynomials of $x$ alone this is obvious since $Q(x)$ is the
charecteristic polynomial of the matrix $x$
and all higher powers of $x$ can be expressed
 as combinations of the above basis modulo $Q(x)$ alone.
Let us show that $y$ can also be expressed as a combination of
polynomials from the above basis.  It is enough to show that it can
be expressed as a general
polynomial in $x$ modulo the derivatives \constraint\ :

 $$ y = h(x) \ ({\rm modulo\ } Q(x),P'(x)+yQ'(x)) \eqn\yeqh $$
 or equivalently
 that there exist polynomials $f(x,y),g(x,y),h(x)$ such that
 $$ y = h(x) + f(x,y) Q(x) + g(x,y)(P'(x)  + yQ'(x)).  \eqn\eqfory $$
 But, since we chose $Q(x)$ to have no degenerate zeroes, $Q(x)$ and
 $Q'(x)$ have no common divisor, so that there exists a
solution $p(x),q(x)$ to the  equation
$$ p(x)Q'(x) + q(x)Q(x) = 1 \eqn\qqprime $$
 and thus we can take $f(x,y)=yq(x), g(x,y)=p(x)$ and $h(x)=-g(x)P'(x)$
to be the solution to \eqfory. We have thus proven that all polynomials
in $x$ and $y$ can be expressed (modulo the derivatives of the potential)
as linear combinations of the basis elements $1,x,...,x^{N-1}$.
Since the equations are satisfied by the matrices corresponding to
$x$ and $y$, and since
the basis elements have the correct fusion rules, this representation
is indeed a good representation of the desired algebra.

Now , let us continue  to the more
interesting case when neither $x$ nor $y$
have degenerate eigenvalues.
To handle
this case we notice that a linear change of variables from $(x,y)$ to
new variables of the form
 $(\xt = ax + by, \yt = cx + dy)$, which is non-singular (namely,
$ad-bc$ is non-zero), does not change the set of points
where the derivatives
of $V$ (as expressed in terms of the new variables) vanish. Explicitly,
if $V(x,y)$ satisfies
${\partial V \over \partial x} = {\partial V \over \partial y} = 0$
only    at the points $(\lambda_i,\mu_i)$, then $V(\xt, \yt)$
will satisfy
${\partial V \over \partial \xt} = {\partial V \over \partial \yt} = 0 $
only    at the points $(a\lambda_i+b\mu_i, c\lambda_i+d\mu_i)$. All we
need to do, therefore, is to find $a,b,c,d$ that satisfy
$ad-bc \neq 0$ and such that
all points $a\lambda_i+b\mu_i$ are different.
Then we can transform to the variables $\xt=ax+by, \yt=cx+dy$ and build
the potential in exactly the same way as above. The proof that the
representation is faithful works in exactly the same way as above
(working in the transformed
basis $1,ax+by,(ax+by)^2,...,(ax+by)^{N-1}$ instead of
the one above) and can also serve to obtain explicit fusion rules from
the potential in any desired basis (although since we have
used the fusion
rules to build the potential in the first place, this does not give us
additional information).

It should be mentioned, that although the eigenvalues $\lambda_i$ and
$\mu_i$ are in general non-rational, the potential we have
obtained will
have rational coefficients (obviously we can then write it with
integer coefficients as well) as long as the parameters $a,b,c,d$ are
rational. Before the transformation to the new variables,
it is trivial that $Q(x)$ described above has integer
coefficients, since it is the charecteristic polynomial of the integer
valued matrix representing $x$.     The coefficients of $P'(x)$ described
above can also be written in terms of traces and determinants of matrices
of the form $x^ny^m$ which are of course also integers. Thus, $P(x)$
will also have rational coefficients. It is then obvious that if
$a,b,c,d$ are all rational, this property remains true after the
transformation to $(\xt,\yt)$ as well.

As an example, let us analyze the $(4,5)$ minimal model
(the tri-critical Ising model) which was also
analyzed as an example in the previous section, where the eigenvalues
of it's fields were given. From the eigenvalues we can easily see which
pair of fields can generate a two-variable fusion potential
representation.  For example we can choose $x$ to be $\phi_{(1,2)}$ and
$y$ to be $\phi_{(2,1)}$. The simplest choice for the change of variables
is $a=b=d=1,c=0$ so that $\xt = x + y, \yt = y$, and in this
case the polynomials we get from the above procedure are
$$ Q(\xt) = (\xt^4-2\xt^3-5\xt^2+6\xt-1)(\xt^2+\xt-1) \eqn\exaq $$
and
$$ P'(\xt) = -4(\xt^2+\xt-1)(2\xt^2-2\xt-1) \eqn\exap $$
so that the potential turns out to be
$$\eqalign{
 V(x,y) = -{8 \over 5}&\xy^5 + {20 \over 3} \xy^3-2\xy^2-4\xy+ \cr
           y&(\xy^4-2\xy^3-5\xy^2+6\xy-1)\cdot \cr
            &(\xy^2+\xy-1)\cr}  \eqn\pot $$
with constraints (which are linear combinations of
${\partial V \over \partial x}$ and ${\partial V \over \partial y}$)
of the form
$$ (\xy^4-2\xy^3-5\xy^2+6\xy-1)(\xy^2+\xy-1) = 0 \eqn\cpnst $$
and
$$ \eqalign {-4(\xy^2+\xy-1)(2\xy^2-2\xy-1)&+ \cr
y(4\xy^3-6\xy^2-10\xy+6)(\xy^2+\xy-1)&+ \cr
   y(\xy^4-2\xy^3-5\xy^2+6\xy-1)(2\xy+1)&= 0. \cr } \eqn\stam $$
By the procedure analyzed above, we can represent all fields $x^ny^m$
as linear combinations of $1,\xy,\xy^2,...,\xy^5$ modulo the above
constraints. To work in another basis
(in this case a comfortable basis is $1,x,x^2,x^3,y,xy$) we have
to obtain the
transformation between the two bases which enables us to write any field
in our preferred basis.
In this way we can systematically obtain the constraints associated with
the fusion rules in their more recognizable form (recall
$x=\phi_{(1,2)},y=\phi_{(2,1)}$)
$$ \eqalign {x^4&=3x^2-1\cr x^2y&=xy+y\cr
y^2&=x^3-2x+1\cr } \eqn\minimal $$
where all equalities are satisfied modulo the above constraints. We
should note that the constraints \minimal\ cannot be derived directly
from a potential.

For all minimal models, polynomials in two variables (namely,
$\phi_{(1,2)}$ and $\phi_{(2,1)}$) are sufficient to obtain a potential
description of the fusion rules of the theory, but in other models this
is not necessarily the case. The construction described above can quite
easily be generalized to the case of more than two variables : let us
denote the fields we want to generate the polynomial algebra
 as $x^{(i)}$ for
$i=1,...,n$ and their eigenvalues by $\lambda_j^{(i)}$ for $j=1,...,N$.
Again we may assume that $x^{(1)}$ has no degenerate eigenvalues,
otherwise we can make a linear transformation to new variables in which
that will be the case. For this case a suitable potential is:
$$ \eqalign { V(x^{(i)}) =&P(x^{(1)}) +
x^{(2)}\prod_{i=1}^N (x^{(1)}-\lambda_i^{(1)}) +\cr
&\sum_{l=3}^n \sum_{i=1}^N {1\over 2} (x^{(l)}-\lambda_i^{(l)})^2
\prod_{j\neq i}(x^{(1)}-\lambda_j^{(1)}) \cr } \eqn\genv $$
where $P(x^{(1)})$ is chosen so that the polynomial
${\partial V \over \partial x^{(1)}}$, which is linear in $x^{(2)}$, will
vanish at $x^{(2)}=\lambda_i^{(2)}$ and $x^{(j)}=\lambda_i^{(j)}$
for all other variables. This potential works because the derivative with
respect to $x^{(2)}$ vanishes only when $x^{(1)}$ is at one of it's
eigenvalues, then the derivative with respect to $x^{(l)}$ (for
$l=3,...,n$) vanishes only when $x^{(l)}$ is at it's corresponding
eigenvalue, and the derivative with respect to $x^{(1)}$ forces that
to be the case for $x^{(2)}$ as well. The proofs that this potential
has rational coefficients, and that it gives a faithful representation
of the fusion algebra, work for this case in the same way as in the two
variable case. Since for $n$ large enough such a representation is
available for any fusion algebra, this proves Gepner's conjecture
\refmark\DG.

\chapter{ Summary and conclusions}
In this paper we analyzed various forms of potential representations
of fusion algebras. We started by analyzing the generalized one-variable
representation suggested in ref. [\DFZ], giving a simple criterion to
determine (given the fusion rules) when such a representation is
possible. We have also given a simple way to construct all such
representations (there is always a finite number of them).
Representations of this sort, however, do not appear to have a
simple "physical" meaning, since in physical theories we usually allow
fields to be multiplied by any real number and not just by rational
numbers.

We then went on to analyze usual fusion representations in more than
one variable, showing that any theory can be represented in such a way.
In fact, we have shown that whenever there is any polynomial
representation for an algebra where the variable $x_i$ represent certain
fields $\phi_i$ (which is always available given enough fields as shown
in ref. [\DG]) we can find a potential representation in the same
variables. Unfortunately, the representation of this sort is far from
unique.  Even for a given   choice of
fields as generators of the algebra there is an
infinite number of representations of this sort, corresponding
for example to different solutions of equation \eqforp .
The geometrical meaning of this kind of
 representations was analyzed in [\DG], where the algebra
was interpreted as the algebra of modular transformations of the
hyper-surface $V=0$. Their physical meaning can perhaps be derived from
Landau-Ginzburg models.  For this we would like to interpret them as
being perturbed from some conformal point, with the fields themselves
(which have to be given externally in the simple potential
representation) given as the derivatives of the potential with respect to
the various available perturbations. It is not yet clear to us
when such an
interpretation is possible, and perhaps it can serve to limit the number
of possible potential representations.

I would like to thank Prof. S. Yankielowicz for suggesting to me
the subject of this work and for discussions on this subject.

\refout
\end